\documentclass[a4paper]{jpconf}
\usepackage{grffile}
\usepackage{xcolor}

\usepackage[utf8]{inputenc}
\usepackage{amssymb}
\usepackage{amsmath}
\usepackage{multirow}

\usepackage[pdftex]{graphicx}

\begin{document}
\title{Macroscopic transport of a current-induced spin polarization}

\author{S Ullah$^1$, G J Ferreira$^2$, G M Gusev$^1$, A K Bakarov$^3$ and F G G Hernandez$^1$}

\address{$^1$ Instituto de F\'{i}sica, Universidade de S\~{a}o Paulo, S\~{a}o Paulo 05508-090, SP, Brazil}
\address{$^2$ Instituto de F\'{i}sica, Universidade Federal de Uberl\^{a}ndia, Uberl\^{a}ndia 38400-902, MG, Brazil}
\address{$^3$ Institute of Semiconductor Physics and Novosibirsk State University, Novosibirsk 630090, Russia}

\ead{felixggh@if.usp.br}

\begin{abstract}
Experimental studies of spin transport in a two-dimensional electron gas hosted by a  triple GaAs/AlGaAs quantum well are reported. Using time-resolved Kerr rotation, we observed the precession of the spin polarization about a current-controlled spin-orbit magnetic field. Spatially-resolved imaging showed a large variation of the electron $g$-factor and the drift transport of coherent electron spins over distances exceeding half-millimetre in a direction transverse to the electric field.

\end{abstract}
\section{Introduction}
For the successful implementation of new functionalities using the spin degree of freedom in technological platforms, experimental techniques are required for the generation of spin polarization, transport across relevant length scales and polarization detection. For device applications of non-equilibrium electron spins, triggered by either optical \cite{Saeed2016} or electrical techniques \cite{Felix2014}, the generation and detection of spin currents should be accomplished without the use of extremely strong magnetic fields and desirably using only electrical voltages. The electrical generation of spin polarization have been widely studied in semiconductor structures of different dimensionality such as bulk samples \cite{Kato2004,Stern2006} and in n- and p-doped quantum wells (QWs) \cite{Sih2005,Silov2004}. The transport of spins polarized by an electrical current is of great interest not only due to the emergence of new phenomena but also due to possible applications in spin based electronics \cite{Aronov1989,Edelstein1990}. Transverse transport of in-plane current-induced spin polarization (CISP) have been studied in L-shaped strained n-InGaAs channels \cite{Kato2005}. Furthermore, it has been reported that the spins polarized in the out-of-plane direction by the spin Hall effect (SHE) can be drive long distances in n-doped GaAs epilayers \cite{Sih2006}. In this paper, we report on the electrical generation of spin polarization in two-dimensional electron gases under the action of spin-orbit fields and the spin transport over distances exceeding half millimetre away from current path in a transverse direction.

\section{Materials}
\begin{figure}[h]
\begin{center}
\includegraphics[width=0.65\textwidth]{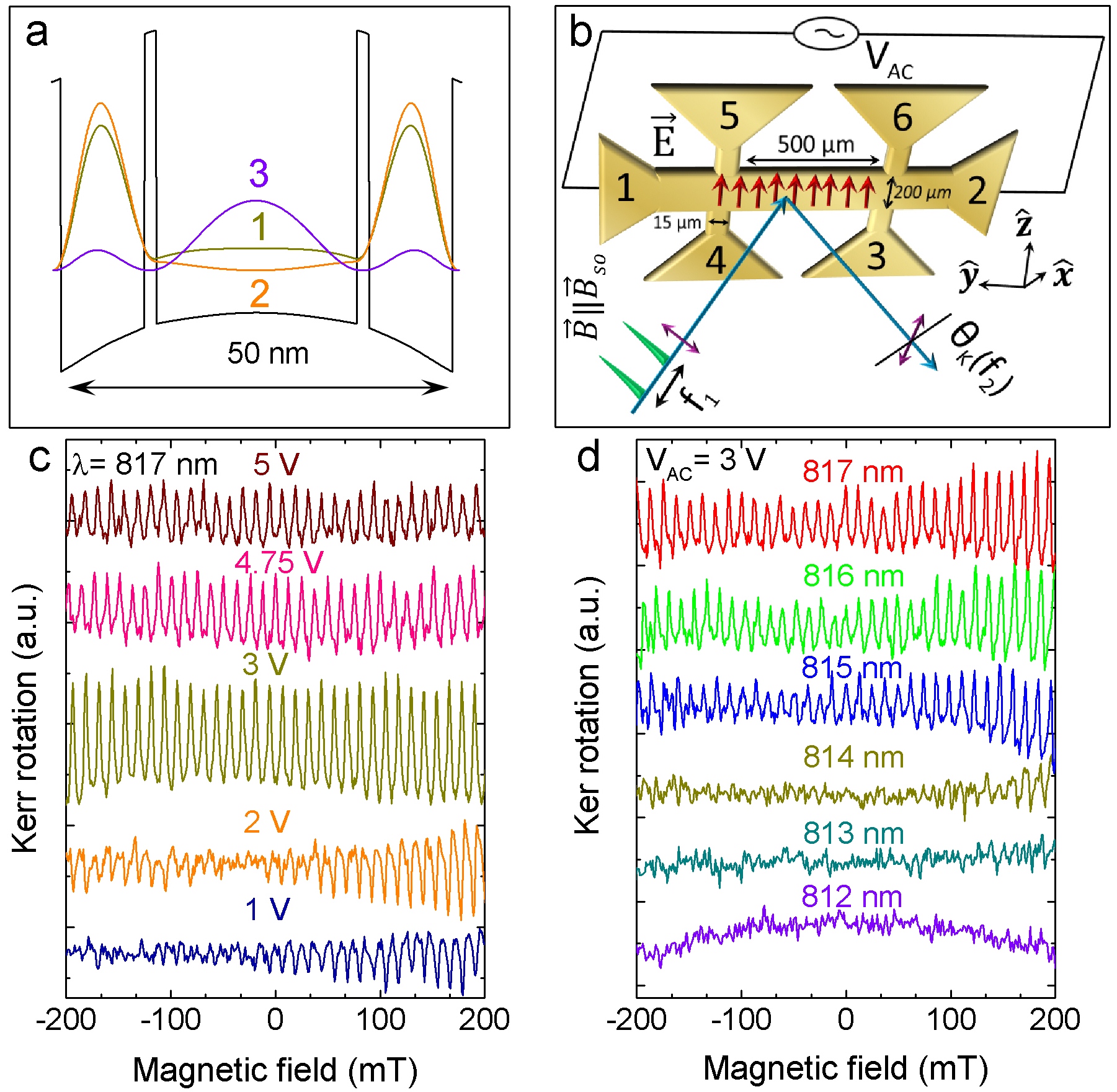}
\end{center}
\caption{Longitudinal configuration: (a) TQW band structure and subband charge density. (b) Schematic for the optical amplification of the CISP. (c) KR versus B measured for several applied voltages. (d) B scan of KR signal at different pump-probe wavelengths. Measurements were taken at a temperature T = 1.2 K and probe power of 300 $\mu$W.}
\label{fig:LCISP}
\end{figure}

The structure discussed in this manuscript is GaAs/AlGaAs triple quantum well (TQW), containg a dense two dimensional electron gas (2DEG) grown in the [001] direction. The structure is symmetrically $ \delta $-doped, with  total density $ n_{s} = 9.6 \times 10^{11} cm^{-2} $ and low temperature mobility $ \mu = 5.5 \times 10^{5} cm^{2}/Vs $. It contains a 26-nm-thick GaAs central well with electron density of about $ 1.4 \times 10^{11} cm^{-2} $ and two 12-nm-thick lateral wells with approximately equal electron density of $ 4.1 \times 10^{11} cm^{-2} $, each separated by 1.4-nm-thick Al$ _{0.3} $Ga$ 0.7 $As barriers. Fig. \ref{fig:LCISP} (a) shows the calculated TQW band structure and subband charge density, where the black lines show the potential profile and the color lines show the first (dark yellow), second (orange) and third (violet) occupied subbands.
For the electrical generation of spin polarization the sample was patterned into a six-contact Hall bar geometry. The main channel have length L = 500 $ \mu m $ between the side probes (in the y-axis), width $W$ = 200 $ \mu m $ and four transverse channels of equal width 15 $ \mu m $ that extend out from the main channel (in the x-axis) as shown schematically in Fig. \ref{fig:LCISP} (b).
\section{Experimental results}
The current-induced spin polarization was measured by using the optical Kerr rotation (KR) technique. The spin polarization was generated by an electrical voltage modulated at $ f_{2} $ = 1.1402 kHz for lock-in detection. A weak linearly polarized probe beam emitted from a pulsed Ti:sapphire laser was scanned on the structure for the spatial detection of the induced spin polarization. The polarization of the reflected beam was measured by a balanced photodetector and lock-in amplifier operating at $f_{2}$. The laser wavelength was tuned for maximum KR signal.
The sample was immersed in the variable temperature insert of a split coil superconductor magnet in the Voigt geometry.

\subsection{Longitudinal spin transport}
In the longitudinal configuration, the KR was measured as a function of external magnetic fields while applying V$ _{AC} $ at the central channel (1-2) [see Fig. \ref{fig:LCISP} (b)]. If the ensemble coherence time ($ T_{2}^{*} $) is longer than the pulse repetition (with MHz frequency) the signal does not fully decay from the previous pulse and overlaps with the signal generated by the following pulses. Due to the constructive interference at certain resonant fields, the amplification of the precession (with GHz frequency) of those electrically polarized spin occurs and result in sharp resonance peaks. These resonance peaks with spacing $\Delta B$ obey the periodic condition $ \Delta B = (hf_{1}/g\mu_{B}) $ where $ f_{1} $ is the laser repetition rate \cite{Felix2014}. Fig. \ref{fig:LCISP} (c) displays the RSA pattern as a function of the external magnetic field for several applied voltages. The CISP amplitude increases with the increase of applied voltage and reaches its maximum value at 3 V. However, further increase of the applied voltage results in a strong reduction of the amplitude possibly due to heating effects \cite{Felix2014}. Fig. \ref{fig:LCISP} (d) displays the dependence of the CISP amplitude on the probe wavelength at a fixed voltage of 3 V. Furthermore, the hole spin coherence time involved in the generation of the electron spin coherence, causes the reinforcement of the resonant peak amplitude at higher magnetic fields \cite{Yugova2009}. The experimental conditions for maximum signal amplitude, namely $V_{AC}$ = 3 V and $\lambda $ = 817 nm, were used in the following section.

\subsection{Transverse spin transport}
\begin{figure}[h]
\begin{center}
\includegraphics[width=0.65\textwidth]{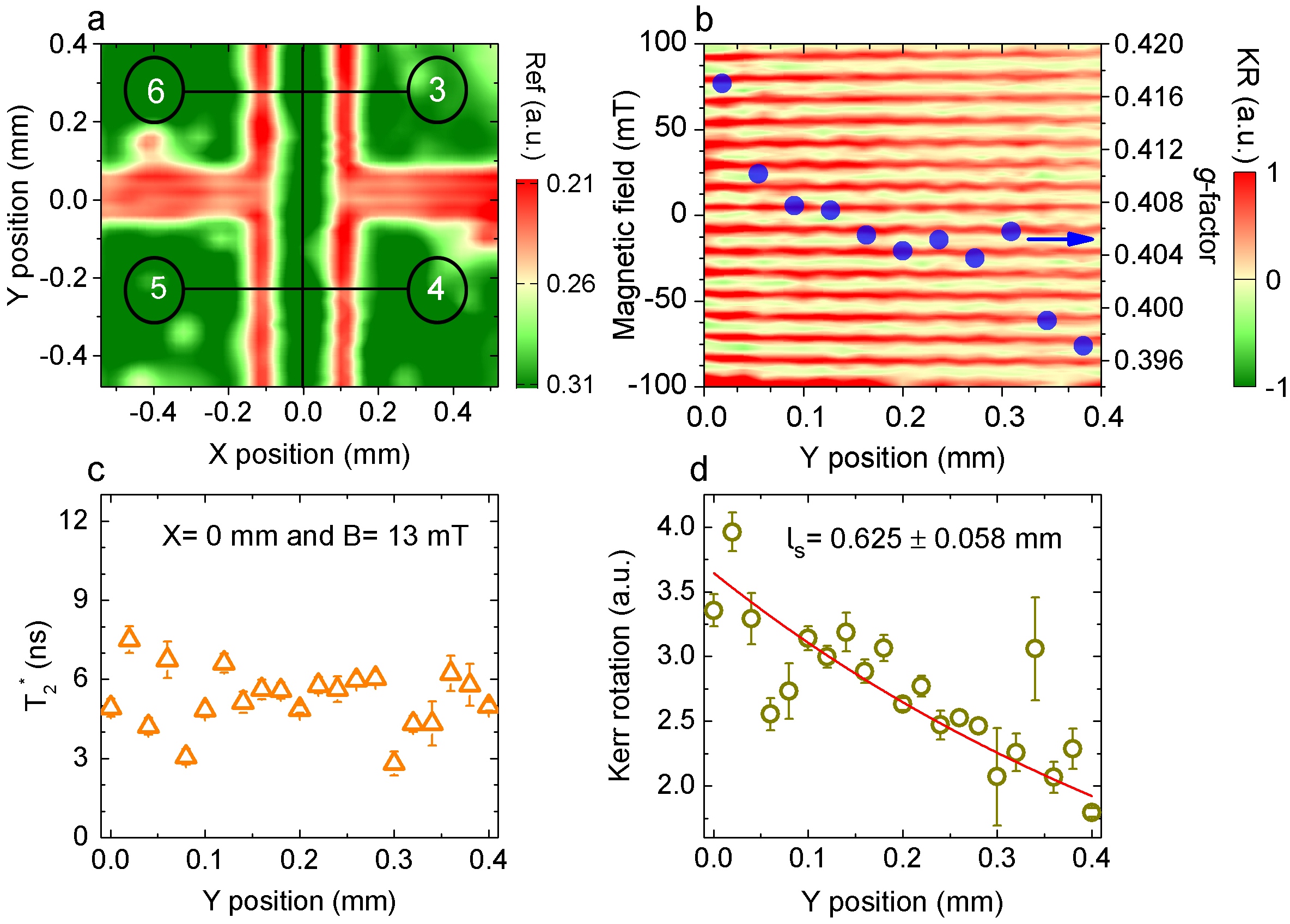}
\end{center}
\caption{\label{fig:TCISP}Transverse configuration: (a) Reflectivity map of the device. (b) KR as a function of B along (0,Y) mm axis with V$ _{AC} $= 3 V. The symbol ($\textcolor{blue}{\fullcircle}$) shows the spatial variation of electron $ g $-factor. The extracted values of $ T_{2}^{*} $ and amplitude are plotted in (c) and (d) respectively, the solid line is an exponential fit to the data.  T = 5 K.}
\end{figure}

Now we switch from to the longitudinal transport to the transverse transport. In contrast to the longitudinal transport, where the spin polarization was measured in a region where the local current flows, we measured the spin polarization in a region with minimal electric field. For example, when $V_{AC}$ is applied along the main channel (y-direction) the transverse transport will be into the side arms (x-direction) of the Hall bar. The reflectivity of the structure imaged as function of the position (x,y) is shown in Fig. \ref{fig:TCISP} (b). The reflectivity map displays the central channel and the four transverse voltage probes. The solid lines highlight the possible current paths and the red regions separate the conducting areas where the white lines are the edges of those regions. In Fig. \ref{fig:TCISP} (b), we imaged the KR signal as a function of B while applying $ V_{AC} $ in the x-axis using contacts 5-4 and scanning the probe spot at (0,y) approximately 250 $ \mu $m away from the current path. We observed the transverse drift of the CISP in a scan of about $ \Delta $y = 0.4 mm. We also observed that the spacing $\Delta$B between the RSA peaks varies with the position. One can clearly see that the outer peaks are shifted towards higher fields for an increasing y-position indicating the variation of the $g$-factor as plotted in Fig. \ref{fig:TCISP} (b). Recently, the $ g $-factor modification in a bulk InGaAs epilayer by an in-plane electric field was studied \cite{Sih2015}. They found a dependence of $\Delta g$ on the drift velocity: $ \Delta g \propto \upsilon_{d}^{2}$. Furthermore, the dependence of the electrical $g$ tensor variation on the magnetic field was also reported in quantum wells \cite{Salis2014}. Our data gives a $g$-factor variation in agreement with those reports.

To extract the spin lifetime and polarization amplitude, the magnetic field scans were fitted using a Lorentzian function $\Theta_{K}= A/[(\omega_{L} T_{2}^{*})^{2} + 1]$ with $ T_{2}^{*}= \hbar/(g \mu_{B} B_{1/2})$, where $\omega_{L}$ is the Larmor frequency, $ \hbar $ is the reduced Planck's constant, $\mu_{B}$ is the Bohr magneton and B$ _{1/2}$ is the peak half-width. The extracted parameters are plotted in Fig. \ref{fig:TCISP} (c) and (d). Surprisingly we found that in our structure the CISP transverse drift can drive a constant $ T_{2}^{*} $ of about 6 ns by almost half millimeter. The transport length of $\ell_{s} $= 0.625 $\pm$ 0.058 mm was extracted from fitting the data in Fig. \ref{fig:TCISP} (d) by $ exp(-y/\ell_{s}) $. Those results of $ T_{2}^{*} $ and $ \ell_{s} $ are independent of the field chosen for analysis. More detailed analysis, theoretical calculations and transport along x-axis are published in \cite{Felix2016}.

\section{Conclusion}
We performed a detailed study on the CISP transport in a two-dimensional electron gas confined in GaAs/AlGaAs quantum wells. We found long $ T_{2}^{*} $ in the nanoseconds range and macroscopic drift distances in a direction transverse to the applied electric field. Additionally, a large spatial variation of the electron $g$-factor was observed.

\section{Acknowledgments}
F.G.G.H. acknowledges financial support from Grant No. 2009/15007-5, 2013/03450-7, 2014/25981-7 and 2015/16191-5 of the S\~{a}o Paulo Research Foundation (FAPESP). S.U. acknowledges TWAS/CNPq for financial support.


\section*{References}
\begin{thebibliography}{9}
\bibitem{Saeed2016}
Ullah S, Gusev G M, Bakarov A K and Hernandez F G G 2016 {\it J. Appl. Phys.} {\bf 119} 215701
\bibitem{Felix2014}
Hernandez F G G, Gusev G M and Bakarov A K 2014 {\it Phys. Rev.} B {\bf 90} 041302(R)
\bibitem{Kato2004}
Kato Y K, Myers R C, Gossard A C and Awschalom D D 2004 {\it Phys. Rev. Lett.} {\bf 93} 176601
\bibitem{Stern2006}
Ster N P, Ghosh S, Xiang G, Zhu M, Samarth N and Awschalom D D 2006 {\it Phys. Rev. Lett.} {\bf 97} 126603
\bibitem{Sih2005}
Sih V, Myers R C, Kato Y K, Lau W H, Gossard A C and Awschalom D D 2005 {\it Nat. Phys.} {\bf 1} 31
\bibitem{Silov2004}
Silov A, Blajnov P, Wolter J, Hey R, Ploog K and Averkiev N 2005 {\it Semiconductors} {\bf 39} 1323
\bibitem{Aronov1989}
Aronov A G and Lyanda-Geller Yu B 1989 {\it JETP. Lett.} {\bf 50} 431
\bibitem{Edelstein1990}
Edelstein V M 1990 {\it Solid state commun.} {\bf 73} 233
\bibitem{Kato2005}
Kato Y K, Myers R C, Gossard A C and Awschalom D D 2005 {\it Appl. Phys. Lett.} {\bf 87} 022503
\bibitem{Sih2006}
Sih V, Lau W H, Myers R C, Horowitz V R, Gossard A C and Awschalom D D 2006 {\it Phys. Rev. Lett.} {\bf 97} 096605
\bibitem{Yugova2009}
Yugova I A, Sokolova A A, Yakovlev D R, Greilich A, Reuter D, Wieck A D and Bayer M 2009 {\it Phys. Rev. Lett.} {\bf 102} 167402
\bibitem{Sih2015}
Luengo-Kovac, Macmahon M, Huang S, Goldman R S and Sih V 2015 {\it Phys. Rev.} B {\bf 91} 201110(R)
\bibitem{Salis2014}
Chen Y S, F\"{a}lt S, Wegscheider and Salis G 2014 {\it Phys. Rev.} B {\bf 90} 121304(R)
\bibitem{Felix2016}
Hernandez F G G, Ullah S, Ferreira G J, Kawahala N M, Gusev G M and Bakarov A K 2016 {\it Phys. Rev.} B {\bf 94} 045305
\end{thebibliography}
\end{document}